\documentclass{article}
\newcommand{\bfr}{\begin{flushright}}
\newcommand{\efr}{\end{flushright}}
 
\begin{document}
\title{The Universe as a Topological Defect in a Higher-Dimensional
Einstein--Yang--Mills Theory
}
\author{Atsushi Nakamula\\
Department of Physics, Tokyo Metropolitan University\\
Setagaya, Tokyo 158, Japan\\
and\\
Kiyoshi Shiraishi\\
Institute for Nuclear Study, University of Tokyo\\
Midoricho, Tanashi, Tokyo 188, Japan
}
\date{Acta Physica Polonica {\bf B21}, No.~1, January 1990, pp.~11--16
}
\maketitle
\begin{abstract}
An interpretation is suggested that a spontaneous compactification of
space-time can be regarded as a topological defect in a
higher-dimensional Einstein-Yang-Mills (EYM) theory. We start with
$D$-dimensional EYM theory in our present analysis. A compactification
leads to a $D-2$ dimensional effective action of Abelian gauge-Higgs
theory. We find a ``vortex'' solution in the effective theory. Our
universe appears to be confined in a center of a ``vortex'', which has
$D-4$ large dimensions. In this paper we show an example with $SU(2)$
symmetry in the original EYM theory, and the resulting solution is
found to be equivalent to the ``instanton-induced compactification''.
The cosmological implication is also mentioned.\\
PACS numbers: 04.60.+n, 12.10.Gq, 98.80.Dr 
\end{abstract}

\bigskip

Recently there has been much interest in the study of
higher-dimensional theories. This is because the candidates for
unified theory can be easily and naturally formulated in higher
dimensions \cite{1, 2}. If the theories with more than four dimension
are
to be taken seriously, a mechanism which brings about dynamical
compactiacation of the extra dimensions is needed. Usually the size
of the extra space is considered to be of order of the Planck length in
order that we cannot observe the extra space experimentally in the
editing laboratory. Various mechanisms of compactification are
proposed in the literature, rasing at variety of ``forces'' to curl up
the extra dimensions \cite{1, 2}.

Rubakov and Shaposhnikov obtained a novel mechanism \cite{3}. They
considered a self-interacting scalar theory in higher dimensions and
discussed the possibility that ``we live inside a domain wall''. It is
implicitly suggested that there are many three-dimensional ``worlds''
in the higher dimensions in their scenario. The possibility of ``many
worlds'' is of great hterest particularly in the cosmological context.
However, their analysis did not include the connection to gravity. lt
is conceivable that no static solution coupled with gravity can be
found and such an energetic structure, which is adequately measured in
the Planck mass, will soon collapse. In addition, the existence of
elementary scalar field with appropriate self-couplings has no
compelling reason supported by unification theories.

On the other hand, it is shown in Ref.~\cite{4} that gauge boson-Higgs
scalar systems are derived from the dimensional reduction of
higher-dimensional Yang-Mills theory. Unfortunately, they consider
the dimensional reduction as a mere device to obtain various
breaking patterns or Higgs mechanism.

Our scenario for compactification is the following: First we consider
that the space-time is partially compactified in higher-dimensional
Einstein-Yang-Mills (EYM) theory. Then we obtain an efreetive
gauge-Higgs theory. Second, we consider a topologlcally non-trivial
structure in the effective theory coupled to Einstein gravity. We
suppose that stable static solutions can be constructed in this way.

Because we start with EYM theory and consider the configuration
with non-trivial topology, it is hopeful to solve the fermion
problem in Kaluza-Klein theory \cite{5}. In fact, a simple example has
been found. Simplest effective (neutral) scalar theory can be derived
from $SU(2)$ gauge theory compactified onto $S^3$, three-dimensional
sphere \cite{6}. There is an exact analytic solution of ``kink'' or
domain wall. This solution turns out to be equivalent to
``instanton-induced compactification'' presented by Randjbar-Daemi,
Salam and Strathdee \cite{7}. Accordingly, the stability has already
been guaranteed in this case.

By imagining the two-step compactification, or ``double
compactification'', we anticipate some kind of phase transition in
the early history of the universe. As usual with Higgs potential, the
effective potential is deferent on temperature and the deformation of
the shape of the potential leads to phase transition \cite{6}. During
the phase transition, some topological defects are formed and we were
born and live inside one of them.

In this paper we adopt $SU(2)$ gauge theory coupled to Einstein gravity
in
$D$ dimensional space-time as an example. First we consider
compactification on $S^2$, two dimensional sphere, and then obtain
effective Abelian Higgs model in $(D-2)$ dimensions. Next we
construct a vortex solution with a unit-quantum of flux. The
classical solution is coupled to gravity. Here the term ``vortex''
means that the topological defect has $(D-5)$ dimensional spatial
extension in $(D-3)$ dimensional space; the extension is two dimension
less than the space it can move.

In the following analysis, we set $D=8$ in practice. The
dimensionality of the resulting flat space-time we live in can be taken
as an arbitrary number, but in this paper the dinensionality of the
flat space-time is set to four.

We begin with the followhg Einstein-Yang-Mills action
\begin{equation}
S=\int
d^8x\sqrt{-g}\left(-\frac{1}{2\kappa^2}R+\frac{1}{4e^2}{\rm
tr}(F_{MN}F^{MN})+\Lambda\right)\,.  
\label{eq1}
\end{equation}
Here $\kappa^2=8\pi G$; $G$ is Newton constant; $e$ is a gauge coupling
constant; $\Lambda$ is a cosmologlcal constant. The scalar curvature of
$S^N$ with unit radius is defined as $R =+N(N-1)$. The suffices $M$ and
$N$ run from $0$ to $7$.

The gauge symmetry group under consideration is $SU(2)$, This may be
regarded as a subgroup of a large unified symmetry group.

The field equations are Yang-Mills equations
\begin{equation}
D_MF^{MN}=\nabla_MF^{MN}+i[A_M, F^{MN}]=0\,,
\label{eq2}
\end{equation}
where the field strength is given by
\begin{equation}
F_{MN}=\partial_MA_N-\partial_NA_M+i[A_M,A_N]\,,
\label{eq3}
\end{equation}
and Einstein equations
\begin{equation}
R_{MN}=\frac{\kappa^2\Lambda}{3}g_{MN}+\kappa^2\left(T_{MN}-
\frac{1}{6}Tg_{MN}\right)\,,
\label{eq4}
\end{equation}
where
\begin{equation}
T_{MN}=\frac{1}{e^2}{\rm
tr}\left(F_{MN}F^P{}_{N}-\frac{1}{4}F_{PQ}F^{PQ}g_{MN}\right)\,,
\end{equation}
and
\begin{equation}
T=T_M^M\,.
\label{eq6}
\end{equation}
Here $R_{MN}$ is the Ricci tensor derived from the metric $g_{MN}$.

To solve the equations coupled to gravity, wetake an ansatz for the
form of the metric:
\begin{equation}
ds^2
=ds^2(M_4)+g^2(\rho)d\rho^2+a^2(\rho)d\psi^2+b^2(\rho)d\Omega^2(S^2)\,,      
\label{eq7}
\end{equation}
where $d\Omega^2(S^2)=d\theta^2+\sin^2\theta d\phi^2$ and
$0\le\psi<2\pi$, $0\le\theta<\pi$ and
$0\le\phi<2\pi$. We assume that the four dimensional space-time we
live in admits flat Minkowski metric.

The form of gauge fields on $S^2$ is assumed to be
\begin{eqnarray}
A_\theta&=&\Phi_1\frac{1}{2}\left(\begin{array}{cc}
0 & -i e^{-i\phi}\\ i e^{i\phi} & 0\end{array}\right)+
\Phi_2\frac{1}{2}\left(\begin{array}{cc}
0 & e^{-i\phi}\\ e^{i\phi} & 0\end{array}\right)\,, \\
\label{eq8a}
A_\phi&=&-\Phi_1\frac{1}{2}\left(\begin{array}{cc}
0 & e^{-i\phi}\\ e^{i\phi} & 0\end{array}\right)+
\Phi_2\frac{1}{2}\left(\begin{array}{cc}
0 & -i e^{-i\phi}\\ i e^{i\phi} & 0\end{array}\right)\nonumber \\
& &+
\frac{1}{2}\left(\begin{array}{cc}
1-\cos\theta & 0\\ 0 & -(1-\cos\theta)\end{array}\right)\,.
\label{eq8b}
\end{eqnarray}

Here $\Phi_1$ and $\Phi_2$ are functions of the coordinates but
independent of the $S^2$ coordinates $\theta$ and $\phi$. Further, we
assume the ``$U(1)$ gauge field'' in six dimensional space-time, in
general, of the form:
\begin{equation}
A_\mu=A_\mu(x^\mu)\frac{1}{2}\left(\begin{array}{cc}
1 & 0\\ 0 & -1\end{array}\right)\,,
\label{eq9}
\end{equation}
where $A_\mu(x^\mu)$ depends only on the coordinate of six dimensions.
$\mu$ runs from $0$ to $5$. Note that here we use a coordinate basis
associated with the metric and not an orthorgonal one.

When the ansatz for the form of gauge configurations is substituted to
the Yang-Mills equation (\ref{eq2}), the equations of motion for
$\Phi_1$, $\Phi_2$ and $A_\mu$ closely resemble the equations of notion
in Abelian Higgs model considered by Nielsen and Olesen \cite{8}. The
equations of motion are
\begin{eqnarray}
& &D^\mu
D_\mu\hat{\Phi}+\frac{1}{b^2}(1-|\hat{\Phi}|^2)\hat{\Phi}=0\,,\\
\label{eq10a}  
& &
\nabla_\mu(b^2F^{\mu\nu})+i(\hat{\Phi}^*D^\nu\hat{\Phi}-\hat{\Phi}D^\nu
\hat{\Phi}^*)=0\,,
\label{eq10b}  
\end{eqnarray}
where $\hat{\Phi}=\Phi_1+i\Phi_2$ and $F_{\mu\nu}=\nabla_\mu
A_\nu-\nabla_\nu A_\mu$. The covariant derivative is defined as 
$D_\mu=\nabla_\mu+iA_\mu$ where $\nabla_\mu$ is the covariant
derivative associated with six-dimensional metric $g_{\mu\nu}$.

Of course the equations coincide with theirs when $g^2=1$, $a^2=\rho^2$
and $b^2$=constant. When gravity is coupled, the solution obtained by
Nielsen and Olesen is modified except for near the origin, $\rho=0$. We
impose an ansatz for the form of a vortex solution. They are the
following \cite{8}:
\begin{eqnarray}
\hat{\Phi}&=&|\Phi|(\rho) e^{i\psi}\,,\\
\label{eq11a}
 A_\mu&=&0\quad\mbox{except for~} A_\psi(\rho) \mbox{~(function on~}
\rho\mbox{)}\,.
\end{eqnarray}

Furthermore we assume physically plausible properties to solve the
equation: In the limit $\rho\rightarrow\infty$, $|\Phi|\rightarrow 1$
and $b\rightarrow 0$, and at $\rho\rightarrow 0$, $g^2=1$,
$a^2\rightarrow\rho^2$ (usual cylindrical coordinates) and the
``magnetic flux'' $\int A_\psi d\psi\rightarrow  0$.

At last, we find the following solution of vortex-type:
\begin{eqnarray}
|\Phi|(\rho)&=&\frac{\rho/B}{\sqrt{1+(\rho/B)^2}}\,,\\
\label{eq12a}
A_\psi(\rho)&=&\frac{1}{\sqrt{1+(\rho/B)^2}}-1\,,\\
\label{eq12b}
g^2(\rho)&=&\frac{1}{({1+(\rho/B)^2})^2}\,,\\
\label{eq12c}
a^2(\rho)&=&\frac{\rho^2}{{1+(\rho/B)^2}}\,,\\
\label{eq12d}
b^2(\rho)&=&\frac{B^2}{{1+(\rho/B)^2}}\,,
\label{eq12e}
\end{eqnarray}
where $B^2=\kappa^2/(2e^2)$ and the natness of the large dimensions
realizes provided that $\Lambda=6e^2/\kappa^4$. These algebraic
relations remain unchanged when the dimensionality of the flat
space-time left untouched varies.

The flux is quantized as expected:
\begin{equation}
\lim_{\rho\rightarrow\infty}\frac{1}{2\pi}\int_0^{2\pi}A_\psi
d\psi=-1\,.
\label{eq13}
\end{equation}
Note that here we obtain an analytic expression of the solution.

For the natural values of the couplings, $B$ is small oforder of the
Planck length. Thus the size or the core of the vortex is nearly the
Planck scale. The direction of $\rho$ is contructed by gravitational
effect, and then it becomes ``compact''. At first sight the metric of
the solution is a little bizarre, but the geometry of the space spanned
by $\rho$, $\psi$, $\theta$ and $\phi$ turns out to be one of $S^4$
after the rewriting of the coordinates.
  
Next, we examine the energy of the configuration of Yang-Mills fields.
  
One can find
\begin{eqnarray}
F_{\rho\psi}&=&-\frac{\rho/B^2}{({1+(\rho/B)^2})^{3/2}}
\frac{1}{2}\left(\begin{array}{cc}1&0\\0&-1\end{array}\right)\,,\\
\label{eq14a}
F_{\theta\phi}&=&\frac{\sin\theta}{{1+(\rho/B)^2}}
\frac{1}{2}\left(\begin{array}{cc}1&0\\0&-1\end{array}\right)\,,\\
\label{eq14b}
F_{\rho\theta}&=&\frac{1/B}{({1+(\rho/B)^2})^{3/2}}
\frac{1}{2}\left(\begin{array}{cc}0&-i
e^{-i(\phi-\psi)}\\i
e^{-i(\phi-\psi)}&0\end{array}\right)\,,\\
\label{eq14c}
F_{\psi\phi}&=&\frac{\rho\sin\theta/B}{{1+(\rho/B)^2}}
\frac{1}{2}\left(\begin{array}{cc}0&-i
e^{-i(\phi-\psi)}\\i
e^{-i(\phi-\psi)}&0\end{array}\right)\,,\\
\label{eq14d}
F_{\rho\phi}&=&-\frac{\sin\theta/B}{({1+(\rho/B)^2})^{3/2}}
\frac{1}{2}\left(\begin{array}{cc}0&
e^{-i(\phi-\psi)}\\
e^{-i(\phi-\psi)}&0\end{array}\right)\,,\\
\label{eq14e}
F_{\psi\theta}&=&\frac{\rho/B}{{1+(\rho/B)^2}}\frac{1}{2}\left(\begin{array}{cc}0&
e^{-i(\phi-\psi)}\\
e^{-i(\phi-\psi)}&0\end{array}\right)\,.
\label{eq14f}
\end{eqnarray}

If one utilizes the vierbeins to treat the suffices, one can
immediately see the anti-self duality of the solution. Thus the energy
density per unit three-dimensional large spatial volume is given by
\begin{equation}
\frac{1}{4e^2}\int d^4y\sqrt{g(y)}{\rm tr~} F^2=\frac{1}{4e^2}
\left|\int d^4y{\rm tr~} F\tilde{F}\right|=\frac{4\pi^2}{e^2}\,,
\end{equation}
where $y^m=\{\rho, \psi, \theta, \phi\}$ and $g(y)={\det} g_{mn}$.
  
The stability of the solution is expected from this relation. though the
full stability analysis should include the perturbation of the metric
and the mixing with the modes from graviton. The relation in the form of
the gauge configuration exhibits the equivalence of the well-known
solution to the one of the ``instanton-induced compactification''
\cite{7}. We emphasize, however, that when the technique of the type we
showed is applied to an EYM theory with other gauge group and
compactification, sub as $SU(m+1)$ on CP$^m$ ($m\ge 2$) \cite{4} we can
types of solutions. We will not explore it further here.

In the model of the present type, we cannot take arbitrary values for
coupling constants of the scalar $\Phi$. It can be read from the
equation of motion that the effective ``Higgs self-coupling'' is the
same as the gauge coupling up to an adequate normalization \cite{9}.
This fact suggests the existence of a static multi-vortex solution or a
configuration of $n$-super-imposed vortices. We study of multi-vortex
system is the most important task when wc wish to investigate a
cosmological scenario (see later).

To conclude this paper, we should mention the cosmological
implication. As mentioned before, we wish to consider ``double
compactification'' as a dynamical phase transition, not merely a
technical formulation of compactification. The investigation of the
high-temperature phase of the model and the study of dynamical
evolution of the scale factors are of great importance if one wants to
consider inflation  or cosmological aspects.

Another remarkable concept is the possibility of ``many worlds''. If
topological defects can be copiously produced after the phase
transition, many lower dimensional ``worlds'' can form networks. The
event of ``worlds in collision'' may occur in higher dimensional
space. In analogy with cosmic strings \cite{10}, self-crossing of a
vortex as a ``world'' may lead to a ``closed'' world. The questions now
arise: does it then collapse? If we live in the vortex, how about the
effect of the collision or crossing on our world? The above questions
must be considered not only by analysing classical solutions of
multi-defect system but also by including quantum effects of matter
and gravity. The selection or the manifold on which the first-step
compactification produces the gauge-Higgs system may also be influenced
by quantum effects.

As a variation of the scenario, one can consider that we live outside
the defects and then we can find the extra space only inside the
topological defect in ow universe. This possibility is also worth
studying.

If the solution obtained here is included in some series of
solutions, we must find the method to obtain the series of the
solutions and eitend the solutions to apply to various physical
situations. We want to investigate other gauge configuration in other
type of the compact manifolds in future works.

\bigskip

\bigskip

One of the authors (K.S.) would like to thank S.~Saito and H.~Yamakoshi
for some coments.

This work is supported in part by the Grant-in-Aid for Encouragement of
Young Scientist from the Ministry of Education, Science and Culture (\#1
6379015O).

One of the authors (K.S.) is grateful to the Japan Society for the
Promotion of Science for the fellowship. We also thanks Iwanami
F\=ujukai for financial aid.


\end{document}